\RequirePackage{fix-cm}
\documentclass[smallextended]{svjour3}       
\smartqed  
\usepackage{graphicx}
\usepackage{multirow}
\usepackage{tabu}
\usepackage{amssymb}
\usepackage{chngpage}
\usepackage{adjustbox}

\begin{document}

\title{Increasing Validity Through Replication: An Illustrative TDD Case}

\author{Adrian Santos * \and Sira Vegas \and Fernando Uyaguari \and Oscar Dieste \and Burak Turhan \and Natalia Juristo 
}


\institute{* Adrian Santos, corresponding author \at
              M3S-ITEE University of Oulu, Finland \\
              \email{adrian.santos.parrilla@oulu.fi}           
           \and
           Sira Vegas, Oscar Dieste, Natalia Juristo \at
              Escuela T\'ecnica Superior de Ingenieros Inform\'aticos, Universidad Polit\'ecnica de Madrid, Spain \\
              \email{svegas/odieste/natalia@fi.upm.es} 
            \and 
            Fernando Uyaguari \at
            ETAPA, Ecuador\\
            \email{fuyaguari01@gmail.com} 
            \and
            Burak Turhan \at Department of Computer Science, Monash University, Australia \\
            \email{burak.turhan@monash.edu}
}

\date{Received: 01/03/18 / Accepted: -}

\maketitle

\begin{abstract}
\textbf{Context:} Software Engineering (SE) experiments suffer from threats to validity that may impact their results. Replication allows researchers building on top of previous experiments' weaknesses and increasing the reliability of the findings. \textbf{Objective:} Illustrating the benefits of replication to increase the reliability of the findings and uncover moderator variables. \textbf{Method:} We replicate an experiment on Test-Driven-Development (TDD) and address some of its threats to validity and those of a previous replication. We compare the replications' results and hypothesize on plausible moderators impacting results. \textbf{Results:} Differences across TDD replications' results might be due to the operationalization of the response variables, the allocation of subjects to treatments, the allowance to work outside the laboratory, the provision of stubs, or the task. \textbf{Conclusion:} Replications allow examining the robustness of the findings, hypothesizing on plausible moderators influencing results, and strengthening the evidence obtained.
\end{abstract}

\section{Introduction}
\label{one}

Isolated experiments are being run in Software Engineering (SE) to assess the performance of different treatments (i.e., tools, technologies or processes). However, isolated experiments suffer from several shortcomings \cite{gomez2014understanding}: (1) results may be imprecise (as the number of participants is typically low in SE experiments); (2) results might be artifactual (e.g., influenced by the programming environment rather than the treatments themselves); (3) results cannot be generalized to different contexts rather than those of the experiment; (4) results may be impacted by the materialization of unforeseen threats to validity (e.g., rivalry threat, when the effectiveness of the less "desirable" treatment gets penalized by the participants' disinterest).

Replication of experiments may help to overcome such limitations \cite{menzies2016perspectives}. For example, replications' designs can be tweaked to overcome the threats to validity of previous experiments. This allows comparing the replications' results and hypothesizing on whether the replications' threats to validity might have materialized or not. Besides, replications allow the systematic variation of specific elements from baseline experiments' configurations (e.g., the experimental task). This allows studying the effect of such changes on results, and thus, elicit moderator variables (i.e., variables rather than the treatments impacting the results \cite{juristo2009using}). 

As an illustrative example, here we show how we run a replication that overcame the weaknesses of a series of experiments on Test-Driven-Development (TDD) \cite{beck2003test}. TDD is an agile software development process that enforces the creation of software systems by means of small testing-coding-refactoring cycles \cite{beck2003test}. TDD advocates have attributed it several benefits, being the most claimed its ability to deliver high quality\footnote{Along this article we focus on external quality as it is the most researched quality attribute across the literature on TDD \cite{rafique2013effects}. We will use the terms "external quality" and "quality" interchangeably along the rest of the article.} software \cite{beck2003test}. This is because, according to TDD advocates, TDD's continuous in-built testing and refactoring cycles create an ever growing safety net (i.e., a test bed) on which software systems rest \cite{beck2003test}. Learning about TDD's effectiveness in terms of quality may assist practitioners when deciding which development approach to use in their daily practice. 

Specifically, we run a replication of Erdogmus et al.'s experiment on TDD \cite{erdogmus2005effectiveness}. Erdogmus et al.'s experiment has already been replicated by Fucci et al. \cite{fucci2013replicated} retaining the same experimental design, but changing the operationalization of the response variable (i.e., quality). Even though it is not possible to rule out all the validity threats in any experiment \cite{shull2002replicating}, here we illustrate how it is possible increasing the robustness of the findings by running replications that overcome previous experiments' threats to validity.

Specifically, by means of an illustrative case here we show how to: 
\begin{itemize}
    \item{Build on top of previous results and increase the reliability of the findings of a series of replications---by addressing several threats to validity of earlier experiments.}
    \item{Assess whether potential threats to validity may have materialized in previous experiments, and observe the stability of the findings.}
    \item{Identify moderator variables and propose further lines of research based on such findings.}
\end{itemize}

This paper is organized as follows. In Section \ref{two} we show the related work on replication in SE, and also the related work on the effectiveness of TDD in terms of quality. In Section \ref{three} we provide an overview of the characteristics and results of the baseline experiment and its close replication. In Section \ref{four} we report the characteristics and results of the replication on TDD that we run. In Section \ref{ten} we compare the results and settings of the experiments and uncover potential moderators. Finally, Section \ref{six} concludes.

\section{Related Work}
\label{two}

First, we show the related work on replication (see Section \ref{related_work_replication}). Then, we show the related work on the empirical studies on TDD (see Section \ref{related_work_empirical}).

\subsection{Replication in SE}
\label{related_work_replication}
The role of replication in SE has been largely argued \cite{gomez2014understanding}\cite{kitchenham2008role}\cite{de2015investigations}\cite{da2014replication}\cite{bezerra2015replication}. Phrases such as \textit{"...replication is the repetition of an experiment to double-check its results..."} \cite{juristo2009using}, or \textit{"...a replication is a study that is run,..., whose goal is to either verify or broaden the applicability of the results of the initial study..."} \cite{shull2004knowledge} are common in the literature. Summarizing, replications are traditionally seen in SE as a way of either increasing the reliability of the original findings, or generalizing baseline experiments' results to different contexts.

Different types of replications may serve better than others for certain purposes \cite{gomez2014understanding}\cite{kitchenham2008role}. For example, \textit{identical replications} \cite{gomez2014understanding} may serve for verifying previous experiments' results. On their side, \textit{conceptual replications} (i.e., replications only sharing baseline experiment's research questions and objectives) may serve for demonstrating \textit{"that an effect is robust to changes with subjects, settings, and materials..."} \cite{kitchenham2008role}. 

Even though many classifications have been proposed in SE to categorize the different types of replications that can be run \cite{gomez2014understanding}, Gomez et al. \cite{gomez2014understanding} propose to classify replications according to the the dimensions that they changed with regards to baseline experiments:
\begin{itemize}
    \item{\textbf{Operationalization}: refers to the operationalization of the treatments, metrics and measurement procedures (e.g., response variables, test cases, etc.). } 
    \item{\textbf{Population}: refers to the participants' characteristics (e.g., skills and backgrounds, etc.).}
    \item{\textbf{Protocol}: refers to the "apparatus, materials, experimental objects, forms and procedures" used (e.g., tasks, session length, etc.).}
    \item{\textbf{Experimenters}: refers to the personnel involved (e.g., the trainer, the analyst, etc.). }
\end{itemize}

We will use such classification to categorize the replication on TDD we report along this article. 

\subsection{Empirical studies on TDD}
\label{related_work_empirical}
Empirical studies on TDD have studied different quality attributes \cite{rafique2013effects}\cite{munir2014considering}\cite{shull2010we}. However external quality seems the most researched so far \cite{rafique2013effects}\cite{munir2014considering}\cite{shull2010we}. External quality is typically measured by means of acceptance test cases. The larger the number of acceptance test cases passed by an application, the larger its external quality \cite{rafique2013effects}.

Munir et al. classified the empirical studies on TDD according to their rigour and relevance \cite{munir2014considering}. Rigour was defined as the adherence to good research and reporting practices. Relevance as the practical impact and realism of the setup. Based on rigor and relevance, Munir et al. mapped the studies on TDD into a 2x2 grid classification (high rigour/high relevance, etc.) \cite{munir2014considering}. The nine studies in the high rigour and relevance category (industry and academic case studies and surveys) show improvements in external quality. Studies with high relevance and low rigour (industry and academic case studies) obtained similar results. Studies with low relevance and high rigour (experiments and one case study with students) mostly show no differences in terms of external quality between TDD and other development approaches. The results of studies with low relevance and rigour (experiments, a survey and a case study with students) show both positive and neutral effects for TDD. 

On their side, Shull et al. classified the studies on TDD in three categories \cite{shull2010we}: (1) controlled experiments; (2) pilot studies or small studies in industry; and (3) industry studies (i.e., studies on real projects under commercial pressures). After classifying the studies, Shull et al. claim that there is moderate evidence that TDD tends to improve external quality \cite{shull2010we}. However they note that the inconclusive---and at times contrary---results reached in the studies might arise due to the different constructs used to measure the response variables, the different control treatments (waterfall, iterative test last, etc.), differences across environments, or the varying participants' expertise. 

The above reviews include different types of empirical studies on TDD (surveys, case studies, experiments etc.). This might have had an impact on the conclusions reached. However, a systematic literature review performed by Rafique and Misic \cite{rafique2013effects}, including only experiments on TDD---and a series of meta-analyses \cite{borenstein2011introduction}---show that TDD appears to result in quality improvements (being such improvement much larger in industry than in academia \cite{rafique2013effects}). Also, Rafique and Misic mention that differences across experiments' results might be due to the different control treatments applied (Waterfall, iterative test last, etc.). 

These observations seem to point towards the necessity of running experiments on TDD relying on similar technological environments, participant expertise, control treatments, and response variables' operationalizations. A systematic approach towards replication on TDD's experiments may strengthen the reliability of the findings, and shed light on plausible reasons for the contradictory results observed in the literature.

\section{Baseline Experiment and Previous Replication}
\label{three}

Along this section we provide an overview of the settings and results of the baseline experiment on TDD (i.e., Erdogmus et al. \cite{erdogmus2005effectiveness}, \textit{BE} onwards), and its close replication (i.e., Fucci et al.  \cite{fucci2013replicated}, \textit{RE1} onwards). We follow Carver's guidelines for reporting experimental replications \cite{carver2010towards}.

\subsection{Research Question and Response Variable}
\label{research_questions}

The independent variable across the replications is \textbf{development approach} (with TDD or ITL as the treatments). Both TDD and ITL follow the same iterative steps (i.e., decomposing the specification into small programming tasks, coding, test generation, execution, and refactoring). The only difference is the order in which the tests are created: before coding in TDD and after coding in ITL. 

\textbf{BE} and \textbf{RE1} share the same research question:
\begin{itemize}
    \item\textbf{RQ}: Do programmers produce \textit{higher quality} programs with TDD than with ITL? 
\end{itemize}
    
The same response variable is studied in BE and RE1: \textbf{quality}. The calculation of quality in BE and RE1 is based on "user stories". User stories do not follow the traditional notion of XP's user story (i.e. a short, high-level feature scenario written from the perspective of a particular user role, in the language of the user, with a series of success criteria typically tested with black-box acceptance test cases at the application level). Instead, user stories refer to small chunks of functionality that need implementing from a task. User stories are tested in BE and RE1 with user stories' test suites. A user story's test suite contains multiple test cases. A test case is comprised by one or more assert statements. Again, test cases do not follow the traditional notion of XP's acceptance test case (i.e., typically a black box test). Instead, test cases refer to small white box test cases that may overlap with the functionality being tested in traditional acceptance test cases.

As an example, let us imagine a task where the participants have to code some functionality that simulates a robot's behaviour in a multi-dimensional grid. A user story included within such task may be related to the "Movement of the robot along the grid". Test cases such as "test the movement of a robot to the upper left limit of the grid" or "test the movement of the robot to the upper right limit of the grid" may be testing the correct implementation of such user story. The test case to "test the movement of a robot to the upper right limit of the grid" might be comprised by two assert statements:
\begin{enumerate}
    \item{Asserting whether the robot can perform multiple movements up until the upper part of the grid: \\ \texttt{robot.init("(0,0)");\\ robot.move("N",10); \\ assert.equals(robot.getPosition(), "(10,0)")}}.
    \item{Asserting whether the robot can perform movements to the right: \\ \texttt{robot.move("R",10); \\ assert.equals(robot.getPosition(), "(10,10)")}}.
\end{enumerate}

Quality is considered in BE and RE1 as the average quality of the \textit{delivered} functionality. However, slightly different criteria has been considered for measuring QLTY across the replications. Such criteria follows.

\textbf{BE}'s calculation for QLTY requires the calculation of the quality metric of each user story ($QLTY_{i}$). The quality of each user story ($QLTY_{i}$) is given by the percentage of asserts from the user story's test suite that pass. A user story is considered as delivered if at least 50 per cent of the assert statements from such user story's test suite pass. In the computation of QLTY, \textit{only delivered user stories are taken into account}. In particular, QLTY is obtained from averaging all user stories' QLTY ($QLTY_{i}$) by their respective \textit{weight} (being each weight a proxy of the user stories' difficulty based on the total number of asserts in the global test suite). 

\textbf{RE1}'s researchers decided to drop the weights associated to each user story. However, as in BE, RE1's researchers consider a user story as delivered if at least 50 per cent of the user story's tests pass. Overall quality (QLTY) is calculated as the average quality of the delivered user stories. 

Table \ref{metrics_quality} shows a summary of the metrics used across the replications.

\begin{table}[h!]
\begin{center}
\caption{Quality across replications.}
\label{metrics_quality}
{\tabulinesep=1.2mm
\begin{tabu}{ l | l}  \hline
\textbf{ID}& \textbf{Formula} \\ \hline \hline
\multirow{2}{*}{\textbf{BE}} & \(\displaystyle 
QLTY=\frac{\sum_{i=1}^{\#u.s} QLTY_{i}*weight_{i}}{delivered~amount~of~user~stories} 
 \)   \\ 
 
 & \(\displaystyle 
QLTY_{i}=\frac{\sum_{j=1}^{\#asserts} QLTY_{ij}}{\#total~asserts~QLTY_{i}}*100\%, QLTY_{i}>50\% 
 \) \\ \hline
 
\multirow{2}{*}{\textbf{RE1}} & \(\displaystyle 
QLTY=\frac{\sum_{i=1}^{\#u.s} QLTY_{i}}{delivered~amount~of~user~ stories}; \)  \\ 

& \(\displaystyle 
QLTY_{i}=\frac{\sum_{j=1}^{\#asserts} QLTY_{ij}}{\#total~asserts~QLTY_{i}}*100\%, QLTY_{i}>50\% 
 \) \\\hline

\end{tabu}}
\end{center}
\end{table}

\subsection{Participants}

\textbf{BE}'s participants were twenty-four third-year \textit{undergraduate} students taking an eight-week Java course at Politecnico di Torino, Italy. The course was part of the BSc. in Computer Science programme. During the course, all the students learned object oriented and Java programming, basic design concepts (including UML), and unit testing with JUnit.

In \textbf{RE1} a total of 33 \textit{graduate} and 25 \textit{undergraduate students} took part. The participants were from the department of Information Processing Science at the University of Oulu, Finland, and they were participating at a graduate-level course about software quality and testing. The course was comprised by six three-hour lab sessions, where the participants learned about the Eclipse IDE, JUnit, Java, refactoring tools, and TDD. 

\subsection{Design}

\textbf{BE}'s experiment was structured as a 2 groups \textit{between-subjects design}. The treatment group applied TDD, whereas the control group applied ITL. 

BE's participants' skills were measured with a questionnaire delivered before the experiment took place. Afterwards, BE's participants were stratified into three groups attending to their skills (low, medium, high) and then assigned to either ITL or TDD. 13 subjects ended up programming with ITL and 11 with TDD, totalling 24 experimental units.

BE's participants were allowed to work on the task \textit{inside and outside the laboratory}. The participants were only trained on the development approach they needed to apply (either ITL or TDD). Subjects were encouraged to adhere to the assigned development approach as closely as possible during the experiment. When the experiment concluded, the subjects were trained in the other development approach (ITL if they applied TDD, and viceversa).

\textbf{RE1}'s experiment was also set up as a 2 groups \textit{between-subjects design}. Some subjects were grouped in pairs (11 in total) due to space restrictions. Subjects were \textit{randomly} distributed to either ITL or TDD. Four pairs and 16 individuals ended up in the ITL group. Seven pairs and 20 individuals ended up in the TDD group. This totals to 47 experimental units. Subjects were trained in both ITL and TDD before the experimental session. Subjects were allowed to work on the task \textit{only in the laboratory}.

\subsection{Artifacts}

\textbf{BE}'s and \textbf{RE1}'s task was a modified specification of Robert Martin's Bowling Score Keeper (BSK) \cite{martin2001advanced}. BSK was already used in previous experiments on TDD \cite{george2004structured}\cite{munir2014experimental}\cite{fucci2015effects}. BSK's goal is calculating the score of a single bowling game. BSK is algorithm-oriented and greenfield. It does not involve the creation of a graphic user interface (GUI). BSK does not require prior knowledge of bowling scoring rules; this knowledge is embedded in the specification. 

BE's participants were given 8 hours (or more if needed) to finalize the task. Some functionality related to data input and output was dropped from RE1's task to adapt its length to the experimental session time. However, the rest of RE1's task specification was identical to BE's. RE1's participants were given 3 hours to finalize the task. Despite the use of toy tasks---such as BSK---undermines the realism of the settings (at least compared to case studies, where fully fledged applications are typically developed over the course of months with TDD \cite{rafique2013effects}), this allows studying the effectiveness of TDD in a controlled environment. This translates into greater control over external variables that may impact the results (such as the use of different IDEs,  programming languages, etc.). In sum, this translates into sacrificing external validity for the sake of internal validity.

\textbf{BE}'s test suite was comprised by 105 test cases consisting of over 350 asserts. \textbf{RE1}'s test suite was formed by 48 test cases consisting of 56 asserts covering 13 user stories. 

\subsection{Context Variables}

\textbf{BE}'s laboratory was comprised by computers with Internet access, the Eclipse IDE \cite{eclipse2016}, JUnit \cite{massol2003junit}, and an in-built CVS client \cite{cederqvist2002version}. Subjects were \textit{not given a Java stub} to quick-start the experimental task. 

\textbf{RE1}'s laboratory was similar to BE's, but without a CVS client. Besides, a \textit{Java stub was delivered} to the subjects so they could reduce the burden of setting up the environment before beginning the actual task implementation.

\subsection{Summary of Results}

\noindent\fbox{
  \parbox{\textwidth}{
        The difference in quality between TDD and ITL was \textit{not statistically significant} in BE or RE1. However, ITL slightly outperformed TDD in BE, and the opposite happened in RE1.
   }
}

\section{Replication}
\label{four}

Along this section we report the replication on TDD that we run (i.e., \textit{RE2}, onwards). Again, we follow Carver's guidelines for reporting replications \cite{carver2010towards}.

\subsection{Motivation for Conducting the Replication}

RE2's motivation was overcoming the potential limitations posed by BE's and RE1's threats to validity on results. Also, learning the effect of systematic experimental design changes on results. For this, we went over the replications' designs and reflected on their shortcomings and points for improvement. 

For example, BE's and RE1's research question (i.e., Do programmers produce \textit{higher quality} programs with TDD than with ITL?) tests the hypothesis that TDD is superior to ITL. However, to the best of our knowledge, there is no theory stating so. Besides, the evidence in the literature is conflicting (i.e., as positive, negative, and neutral results have been obtained). Thus, to avoid running into a threat to conclusion validity due to the directionality of the effect (as the difference in performance between TDD and ITL with a one-tailed test is more likely to be statistically significant than with a two-tailed test if TDD outperforms ITL \cite{cumming2013understanding}), we removed the directionality of the effect in the research question.  

Asides, BE's subjects were divided into three groups (low, medium, or high) based on their skills before being randomized to either ITL or TDD. This was made to balance out the distribution of the subjects' skills to the treatments. Unfortunately, BE's authors did not report which skills they measured---or how they measured them. Besides, there were no clear cut-off points between the different categories (low, medium, or high), nor a clear definition of how skills' levels were combined to classify a subject within a specific category. RE1 overcame this threat to internal validity by assigning the subjects to the treatments without considering a preliminary set of skills (i.e., by means of full randomization). However, they incurred into another threat to internal validity (i.e., confounding), as some participants were grouped into pairs due to a lack of computer stations. We avoided both threats to validity by making the participants code alone, and apply both treatments twice.

Nuisance factors outside the researchers' control could have affected BE's results---as the participants were allowed to work outside the laboratory. RE1 overcame this threat to internal validity by only allowing the participants to work inside the laboratory. Our replication follows such improvement. 

Another difference across the replications is the maximum allowed time given to the participants to implement the task: eight hours or more in BE, three in RE1 and two hours twenty-five minutes in RE2. By reducing the experimental session's length it is possible avoiding fatigue' potential effect on results. Such time reduction was possible due to the provision of stubs in RE1 and RE2. 

BE's and RE1's thresholds for considering a user story as delivered (i.e., 50 per cent of the user story's assert statements) are artificial and may have impacted quality results. We address this threat to construct validity in RE2 as we set no threshold for measuring quality (see below).

BE's user stories' weighting is another threat to construct validity---as such weights are subjective and dependent upon the test suite implementer. User stories' weighting were removed in RE1 and R2. 

BE's and RE1's participants only code one task. This results in a low external validity of results. In RE2 we improve the external validity of the results by using four tasks. This will allow studying the effectiveness of TDD for different tasks. 

BE's and RE1's participants only apply one treatment. A threat to internal validity named compensatory rivalry threat might have materialized (i.e., loss of motivation due to the application of the less desirable treatment, in this case ITL). RE2's subjects applied both treatments (ITL and TDD) instead of only one. This rules out the possible influence of the compensatory rivalry threat.

RE1's subjects were trained in both TDD and ITL before the experimental session. This might pose a threat to internal validity: leakeage from one development approach to another may materialize if subjects apply a mixed development approach in either the TDD or ITL group. This issue was addressed in BE (as subjects were only trained in their assigned treatment before the experimental session). We also addressed such threat to validity in RE2 by training the subjects only in the treatment to be immediately applied afterwards. 

BE's participants are undergraduate students. This poses a threat to the generalization of the results to other types of developers. RE1's participants are a mixture of undergraduate and graduate students. Although this affords greater external validity, it poses an threat to internal validity due to confounding with subject type. All RE2's participants were graduate students instead of undergraduate students: this may increase the external validity of results.

Table \ref{threats_experiments} shows the threats to validity of the experiments that were overcome in further replications. Rows with (+) represent improvements upon the experimental settings marked with (-). 

\begin{table}[h!]
\begin{adjustwidth}{.99in}{-.99in}  
\small
\begin{center}
\caption{Baseline experiment and replications' threats to validity.}
\label{threats_experiments}
\makebox[\linewidth]{
\begin{tabular}{ l | l | l | l |  l} \hline
\textbf{Validity} & \textbf{Threat} & \textbf{BE}  & \textbf{RE1}  & \textbf{RE2}  \\ \hline \hline
\textbf{Conclusion} &  Directionality & One-tailed (-) &  One-tailed (-) & Two-tailed (+) \\ \hline
\multirow{8}{*}{\textbf{Internal}} 
 & Compensatory rivalry & TDD/ITL (-) &  TDD/ITL (-) & TDD and ITL (+) \\ \cline{2-5}
 & Confounding & No (+) & Pair Programming (-) & No (+) \\ \cline{2-5}
 & Confounding & No (+) & Graduate / Undergraduate (-) & No (+) \\ \cline{2-5}
 & Task Execution & Lab and remote work (-) &  Lab (+) & Lab (+) \\ \cline{2-5}
 & Stubs & No (-) & Yes (+) & Yes (+) \\ \cline{2-5}
 & Task Duration & +8 hours (-) &  3 hours (+) & 2.25h (+) \\ \cline{2-5} 
 & Allocation & Stratified (i.e., skill) (-) &  Full-randomization (+) & Full-randomization (+)\\ \cline{2-5}
& Leakage & No (+) &  Yes (-) & No (+) \\ \hline
\multirow{2}{*}{\textbf{Construct}} &  Operationalization  & 50\% threshold (-)  &  50\% threshold (-) & -  (+) \\ \cline{2-5}
 &  Operationalization & Weighting (-) & None (+)   & None (+)\\ \hline  
\multirow{2}{*}{\textbf{External}} &  Mono-operation bias & BSK (-) &  BSK (-) & BSK, MR, MF, SDK (+) \\ \cline{2-5} 
& Subject type & Undergraduate (-) & Graduate / Undergraduate (-) & Graduate (+) \\ \hline
\end{tabular}}
\end{center}
\end{adjustwidth}
\end{table}

As Table \ref{threats_experiments} shows, RE1 still suffers from threats to validity present in BE. Besides, RE1 also falls into a construct and internal threat that was not present in BE. RE2 overcame validity issues from both experiments \cite{gomez2014understanding}. 

\subsection{Level of Interaction with Original Experimenters}

We used Fucci et al.'s BSK task and its test suite. We had no interaction with Erdogmus et al. 

\subsection{Changes to the Previous Replications}

According to the classification suggested by Gomez et al. \cite{gomez2014understanding}, RE2 modified all BE's and RE1's dimensions:
\begin{itemize}
    \item{\textbf{Opertationalization:} because the response variable's (i.e., quality) operationalization was changed. }
    \item{\textbf{Population:} because the population changed from undergraduate to graduate students. }
    \item{\textbf{Protocol:} because the tasks and session lengths were modified. }
    \item{\textbf{Experimenter:} because the replications were run and analyzed by different researchers. }
\end{itemize}

In the following we provide greater detail on the changes made. 

\subsubsection{Research Question and Response Variable}
We are not aware of any theory indicating that TDD produces higher quality software than ITL. Thus, we removed the directionality of BE's and RE1's RQ. In particular, we restate RE2's RQ as:
\begin{itemize}
    \item\textbf{RQ}: Do programmers produce \textit{equal quality} programs with TDD and ITL?
\end{itemize}

RE2' response variable is \textbf{quality}. We measured quality with acceptance test suites that we (i.e., the experimenters) developed. We used a standardized metric to measure quality: functional correctness. Functional correctness is one of the sub-characteristics of quality defined in ISO 25010 and is described as 'the degree to which a system provides the correct results with the needed degree of precision' \cite{iso25010}. We measure functional correctness as the proportion of passing assert statements ($\#Assert(pass)$) over the total number of assert statements ($\#Assert(All)$). Specifically:

$$QLTY =\frac{\#Assert(Pass)}{\#Assert(All)}*100 $$  

We regard this metric as more straightforward than that used in BE and RE1. First, because it does not require any subjective threshold (e.g., 50\% of assert statements passing to consider a functionality as delivered). Second, because QLTY is no longer bounded between 50 and 100 per cent, and instead, can vary across the whole percentage interval (0\%-100\%). Third, because our metric measures overall quality, and not the quality of the delivered functionality. Thus, subjects delivering smaller amounts of high quality functionality are "penalized" compared to those delivering larger amounts of high quality functionality. 

\subsubsection{Participants}
A one-week seminar on TDD was held at the Universidad Polit\'ecnica de Madrid (UPM) in March 2014. A total of 18 graduate students took part in the seminar. They all had a varying degree of experience in software development and unit testing skills. All subjects were studying for a MSc. in computer science or software engineering at the Universidad Polit\'ecnica de Madrid (UPM). Master's students were free to join the seminar to earn extra credits for their degree programme. The seminar was not graded.

Participants were informed that they were taking part in an experiment, that their data were totally confidential, and that they were free to drop out of the experiment at any time. 

Before the experiment was run the participants filled in a questionnaire. Such questionnaire asked the participants about their previous experience with programming, Java, unit testing, JUnit and TDD. Specifically, subjects were allowed to select one of four experience values: No experience ($<2$ years); Novice ($\ge2$ and $<5$ years); Intermediate ($\ge5$ and $<10$ years); Expert ($\ge$10 years). We code such experience levels with numbers between 1 to 4 (novice,..., expert). Table \ref{subjects_experience} shows RE2's participants' experience levels.

\begin{table}[h!]
\caption{RE2 subjects' experience.}
\label{subjects_experience}
\begin{center}
\begin{tabular}{l|c|c|c|c}
    \hline
    \textbf{Variable} & \textbf{Median} & \textbf{Mode} & \textbf{Min} & \textbf{Max}\\ \hline \hline
    Programming Experience  & 2& 3  & 1 & 3 \\\hline
    Unit Testing Experience  & 1& 1  & 1 & 3 \\\hline
    Java Experience  & 2& 2  & 1 & 3 \\\hline
    JUnit Experience  & 1& 1  & 1 & 2 \\\hline
    TDD Experience  & 1& 1  & 1 & 2 \\\hline
\end{tabular}
\end{center}
\end{table}

As Table \ref{subjects_experience} shows, most of the subjects had five to ten years of experience with programming (mode=3) and from two to five years of experience with Java (mode=2). Besides, the participants had little experience with unit testing or JUnit: fewer than two years (mode=1). Their experience with TDD was also limited (fewer than two years). 

\subsubsection{Design}

\textbf{RE2}'s experiment was structured as a 4 sessions  \textit{within-subjects design}. Within-subjects designs offer advantages over between-subjects designs \cite{brooks1980studying}: (1) reduced variance, and thus, greater statistical power because of the study of within-subjects rather than across-subjects differences; (2) increased number of data points, and thus, greater statistical power as each subject has as many measurements as experimental sessions; (3) subject abilities---over or below the norm---have the same impact on all the treatments (as all subjects are exposed to all the treatments).

RE2's 18 subjects applied ITL and TDD twice in non-consecutive sessions (ITL was applied on the first and third day, whereas TDD on the second and fourth). Thus, up to a total of 18 experimental units multiplied by four sessions (72 experimental units) could be potentially used to study ITL vs. TDD. Subjects were given training according to the order of application of the treatments. Subjects \textit{only worked in the laboratory}.

\subsubsection{Artifacts}

The participants coded four different tasks (i.e., BSK, SDK, MR and MF).

BSK's specifications were reused from Fucci et al. \cite{fucci2013replicated}. Appendix A shows BSK's specification. 

SDK is a greenfield programming exercise that requires the development of various checking rules against a proposed solution for a Sudoku game. Specifically, subjects must deal with string and matrix operations and with embedding such functionalities inside a single API method. We provide the SDK's specifications in Appendix B.

MR is a greenfield programming exercise that requires the development of a public interface for controlling the movement of a fictitious vehicle on a grid with obstacles. MR is a popular exercise used by the agile community to teach and practise unit testing. Appendix C contains MR's specifications. 

MF is an application intended to run on a GPS-enabled, MP3-capable mobile phone. It resembles a real-world system with a three-tier architecture (graphical user interface, business logic, and data access). The system consists of three main components that are created and accessed using the Singleton pattern. Event handling is implemented using the Observer pattern. Subjects were given a description of the legacy code, including existing classes, their APIs, and a diagram of the system architecture (see Appendix D).

Table \ref{number_user_stories} shows the number of user stories, test cases, and asserts for each task's test suite. 

\begin{table}[h!]
\begin{center}
\caption{Number of user stories, test cases, and asserts per task.}
\label{number_user_stories}
\begin{tabular}{ l | l | l | l } \hline
 \multicolumn{2}{c}{\textbf{Task}} & \multicolumn{2}{c}{\textbf{Test suite}} \\ \hline \hline
\textbf{Name} & \textbf{User stories}  & \textbf{Test cases}  & \textbf{Asserts}  \\ \hline
  SDK & 6 & 11 & 13 \\ \hline
  BSK & 13 &  48 & 56 \\ \hline
  MR & 11 &  52 & 89 \\ \hline
  MF & 11 &  45 & 123 \\ \hline
\end{tabular}
\end{center}
\end{table}

\subsubsection{Context Variables}

The experiment was run in a laboratory with computers running a virtual machine (VM) \cite{oracle2015virtualbox} with the Eclipse IDE \cite{eclipse2016}, JUnit \cite{massol2003junit}, and a web browser. Due to time restrictions, subjects received a \textit{Java stub} to help them jump start with the implementation.

\subsection{Analysis Approach}

We run the data analysis with IBM SPSS Statistics Version 24. First, we provide descriptive statistics and box-plots for QLTY. Then, we analyze the data with a Linear Marginal Model (LMM). LMMs are linear models in which he residuals are not assumed to be independent of each other or have constant variance \cite{west2014linear}. Instead, LMMs can accommodate different variances and covariances across time points (i.e., each of the experimental sessions). LMMs require normally distributed residuals. In the absence of normality, data transformations can be used (e.g., Box-Cox transformations \cite{vegas2016crossover}).

Particularly, we fitted a LMM with the following factors: development approach, task, and development approach by task. We included the factor task and its interaction with the development approach to reduce the unexplained variance of the model. After fitting various LMMs with different variance-covariance matrix structures, we selected the unstructured matrix\footnote{With different within-subject variances and covariances across residuals.} as the best fit to the data. This was done according to the criterion of lower 2 log likelihood, and to West et al.'s suggestion \cite{west2014linear}. 

We report the differences in quality across development approaches, tasks and development approaches within tasks. Afterwards, we check the normality assumption of the residuals with the Kolmogorov-Smirnov test and the skewness and kurtosis z-score \cite{field2009discovering}.\footnote{Remember that the skewness and kurtosis values should be zero in a normal distribution. These scores can be converted to z-scores by dividing them by their standard error; if the resulting score is greater than 1.96, the result is significant \cite{field2009discovering}. However, a significant test does not necessarily indicate whether the deviation from normality is enough to bias the statistical procedure applied to the data \cite{field2009discovering}, as the significance level depends heavily on sample size (lower p-values for larger sample size).} Finally, we use QQ-plots to check the residuals' normality assumption.

We complement the statistical results with Hedge's g effect sizes \cite{cook1979quasi}\cite{hedges2014statistical} (Cohen's d small sample size correction  \cite{cohen1977statistical}) and their respective 95 percent confidence intervals (95\% CIs). This may facilitate the incorporation of the results in further meta-analyses \cite{borenstein2011introduction}. We report Hedge's g due to its widespread use in SE \cite{kampenes2007systematic} and its intuitive interpretation: the amount of standard deviations that one experimental group's mean deviates from another. 

\subsection{Data Analysis}
\label{seven}

In this section we show the results of RE2's data analysis. Beware that two subjects dropped from the experiment after the first experimental session, which left us with a drop-out rate of 12.5 per cent. Another two subjects did not deliver any working solution along the experimental sessions, and thus we removed their data from the final dataset. In sum, after pre-processing, a total of 14 subjects made it to the analysis phase. 

\subsubsection{Descriptive Statistics}
\label{prod_itl_tdd}

\begin{figure}[h!]
\caption{QLTY by development approach: box-plot.}
\centering
\label{prod_development_approach}
  \includegraphics[height=7cm]{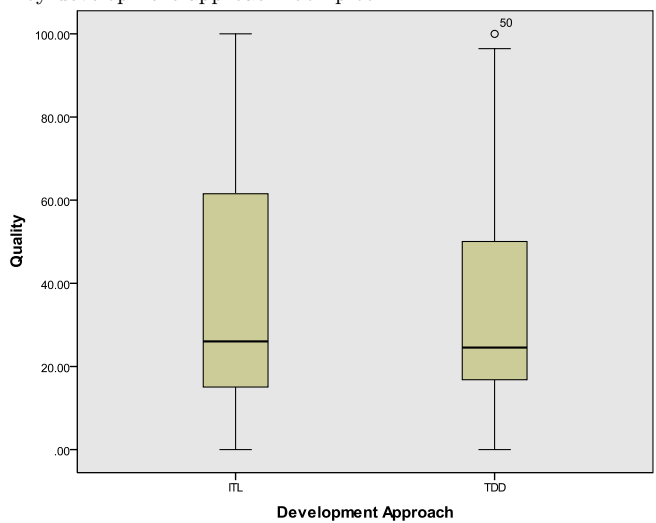}
\end{figure}

Table \ref{descriptives_prod} shows the descriptive statistics for QLTY with TDD and ITL. As Table \ref{descriptives_prod} shows, TDD has a lower mean and standard deviation for QLTY than ITL. This can also be seen in the box-plot for QLTY with TDD and ITL (Figure \ref{prod_development_approach}). However, the 95\% CIs for the means overlap. Thus, \textit{quality seems similar for TDD and ITL}. 

\begin{table}[h!]
\caption{QLTY by development approach: descriptive statistics.}
\label{descriptives_prod}
\begin{center}
\begin{tabular}{l|c|c|c|c}
    \hline
    \textbf{Treatment} & \textbf{N} & \textbf{Mean} & \textbf{95\% CI} & \textbf{SD}\\ \hline \hline
    ITL & 27 & 37.92 & (25.674, 50.161)  & 30.95 \\ \hline
    TDD & 28 & 35.79 & (24.633, 46.962) & 28.792  \\ \hline
\end{tabular}
\end{center}
\end{table}

\begin{figure}[h!]
\caption{QLTY by task and development approach: box-plot.}
\centering
\label{task_prod_by_development_approach}
  \includegraphics[height=7cm]{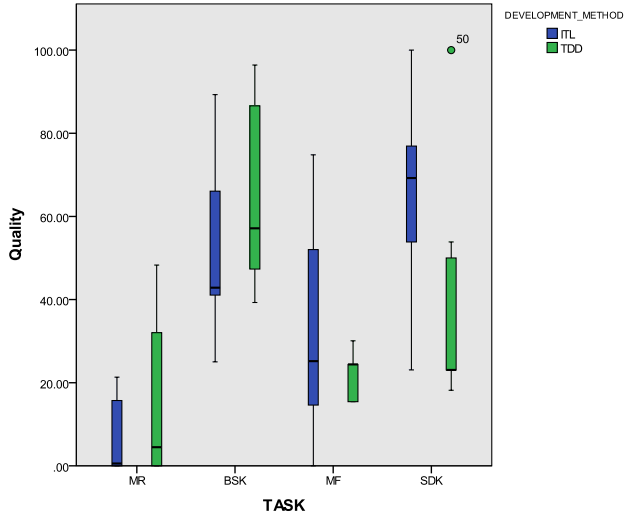}
\end{figure}

Table \ref{descriptives_prod_mr_bsk_mp_sudoku} reports the descriptive statistics for QLTY grouped by task and development approach. Figure \ref{task_prod_by_development_approach} shows the corresponding box-plot. 

As Table \ref{descriptives_prod_mr_bsk_mp_sudoku} shows, MR is the task with the lowest mean QLTY regardless of the development approach. Furthermore, there seems to be a large variability in the QLTY means across development approaches within tasks. As we can see in Table \ref{descriptives_prod_mr_bsk_mp_sudoku}, mean quality within MR varies across development approaches with a 1:2.4 ratio (i.e., 15.168/6.367), whilst such variation seems smaller for BSK, MF and SDK (ratios of 1:1.24, 1:1.47 and 1:1.66 respectively). Furthermore, ratios reverse depending upon the task (i.e., ITL's mean QLTY score is lower than TDD's in MR and BSK, and the opposite in MF and SDK). Thus, an interaction may be taking place.

\begin{table}[h!]
\caption{QLTY by task and development approach: descriptive statistics.}
\label{descriptives_prod_mr_bsk_mp_sudoku}
\begin{center}
\begin{tabular}{c|l|c|c|c|c}
    \hline
    \textbf{Task} & \textbf{Treatment} & \textbf{N} & \textbf{Mean} & \textbf{95\% CI} & \textbf{SD}\\ \hline \hline
    \multirow{2}{*}{MR}& ITL & 6  & 6.367  & (-3.713, 16.446) & 9.605   \\ \cline{2-6}
    &TDD & 8 &  15.168 & (-1.347, 31.684)  & 19.755  \\ \hline
    \multirow{2}{*}{BSK}&ITL & 7 & 53.063  & (29.927, 76.198)  & 25.016    \\ \cline{2-6}
    &TDD & 7 & 65.817 & (44.108, 87.525)  & 23.472    \\ \hline
    \multirow{2}{*}{MF}&ITL & 9 & 32.34  & (11.286, 53.394)  & 27.39  \\ \cline{2-6}
    & TDD & 5 & 21.952  & (14.038, 29.866)  & 6.374     \\ \hline
    \multirow{2}{*}{SDK}&ITL & 5 & 64.616 & (29.137, 100.095)  & 28.57     \\ \cline{2-6}
    &TDD & 8 & 38.812 & (15.507, 62.117)  & 27.876   \\ \hline
\end{tabular}
\end{center}
\end{table}

\subsubsection{Hypothesis Testing}

We fitted a LMM to analyze the data. According to the Kolmogorov-Smirnov test (\textit{p-value=0.007}) and the skewness z-score ($Z_{skewness}=2.10$) the residuals seem to depart from normality. However, the normality Q-Q plot suggests that the residuals follow a normal distribution---despite a few outlying scores at the extremes of the distribution (see Appendix E). 

Due to the stability of the results after applying the Box-Cox transformation, the observation that most of the residuals seem to be following a normal distribution, and the heightened complexity of interpreting the results after data transformations \cite{jorgensen1998obtain}, here we continue interpreting the statistical results of the LMM fitted with the untransformed data. Table \ref{significance_lmm_prod} shows the statistical significance of the factors fitted in the LMM. 

\begin{table}[h!]
\caption{Factor significance in the LMM for QLTY.}
\label{significance_lmm_prod}
\begin{center}
\begin{tabular}{l|l|l}
    \hline
    \textbf{Factor} & \textbf{F} &  \textbf{$p$-value} \\ \hline \hline
    Development Approach & 0.644 & 0.432 \\ \hline
    Task & 53.032 & \textbf{0.000} \\ \hline
    Development Approach*Task & 2.573 & 0.086 \\ \hline
\end{tabular}
\end{center}
\end{table}

As Table \ref{significance_lmm_prod} shows, \textbf{Task has a statistically significant effect} on QLTY at the 0.05 level ($p<0.05$). \textbf{Development Approach*Task has a statistically significant effect} on QLTY at the 0.1 level ($p$-value=0.086). Development approach has not a statistically significant effect on QLTY. As a summary of the results:
\\
\\
\noindent\fbox{
  \parbox{\textwidth}{
            Development approaches seem to behave similarly in terms of QLTY. However, the task under development seems to moderate the effect of the development approach for QLTY. In other words, \textit{the task being developed seems to influence the difference in effectiveness between TDD and ITL}.
   }
}

\subsubsection{Effect sizes}

To ease the comparison of results with BE and RE1 here we report the effect sizes for each task. Table \ref{effect_sizes_task_development_approach_prod} shows that \textbf{effect sizes vary from medium to large across the tasks}. The largest effect size was obtained for SDK, while MF has the smallest effect size (in terms of magnitude). TDD appears to outperform ITL for MR and BSK, whilst the opposite happens for MF and SDK. As a summary, even though none of the effect sizes are statistically significant (as all the 95\% CIs of the effect sizes cross 0), there is an observable heterogeneity of effect sizes across the tasks (as already noticed in the data analysis). 

\begin{table}[h!]
\caption{Hedge's g by tasks for QLTY.}
\label{effect_sizes_task_development_approach_prod}
\begin{center}
\begin{tabular}{l|l|l|l|l} \hline
    \textbf{Task} & \textbf{Effect Size} & \textbf{Hedge's g} & \textbf{95\% CI} & \textbf{Magnitude}\\ \hline \hline
    MR & TDD-ITL	& 0.51  & (-0.508, 1.529)  & Medium \\ \hline
    BSK& TDD-ITL&  0.497 & (-0.51, 1.505)   & Medium \\ \hline
    MF& TDD-ITL& -0.433  & (-1.479, 0.613)   & Medium	\\ \hline
    SDK& TDD-ITL& -0.863  & (-1.96, 0.239)   & Large 	\\ \hline
\end{tabular}
\end{center}
\end{table}

\subsection{Threats to Validity}
\label{nine}
In this section we report RE2's threats to validity following Wohlin et al.'s conventions \cite{wohlin2012experimentation}. We prioritize the threats to validity according to Cook and Campbell's guidelines \cite{cook1979quasi}.

\textbf{Conclusion validity} concerns the statistical analysis of results \cite{wohlin2012experimentation}.

We provide visual and numerical evidence with regard to the validity of the required statistical assumptions. We performed data transformations (i.e., Box-Cox transformations) so as to assess the robustness of the findings. As the results were consistent 
across statistical analyses, for simplicity's sake we interpreted the untransformed data's statistical analysis. Interested readers in the analysis of the transformed data may request them contacting the authors.

The random heterogeneity of the sample threat might have materialized, since the software development experience of the participants ranged from a few months to 10 years. This might have biased the results towards the average performance in both populations, thus resulting in non-significant results.

\textbf{Internal validity} is the extent to which the results are caused by the treatments and not by other variables beyond researchers' control \cite{wohlin2012experimentation}.

A threat to internal validity results from the participants usage of a non-familiar programming environment (e.g. OS and IDE). However, we tried to mitigate this threat by making all the participants use the same environment during the experiment. We expect, thus, that the environment has an equal impact on both treatments---and thus, does not affect results.

There is a potential maturation threat: the course was a five-day intensive course on TDD and contained multiple exercises and laboratories. As a result, factors such as tiredness or inattention might be at work. To minimize this threat we offered the students to choose the schedule that best suited their needs before starting the experiment. We also ensured that subjects were given enough breaks. However, this threat might have materialized due to the drop in quality observed with TDD in the last session (Friday).

Training leakage effect may have distorted results. Even though this training leakage effect was out of the question in the first session (as the subjects were only trained in the development approaches when necessary), it was a possibility in the second, third, and fourth sessions. In particular, subjects might have applied a mixed development approach when they had knowledge of both development approaches. They may have also applied their preferred technique. To mitigate this threat to validity we encouraged subjects to adhere to the development approaches as closely as possible in every experimental session.

There was also the possibility of a diffusion threat: since subjects perform different development tasks in each experimental session, they could compare notes at the end of the sessions. This would give them knowledge in advance about the tasks to code in the following days. This could lead to an improvement in their performance. To mitigate this threat we encouraged subjects not to share any information on the tasks until the end of the five-day training course. Furthermore, we informed the subjects that their performance was not going to have an impact on their grades. Thus, we believe that the participants did not share any information as requested. Since quality dropped in the second application of TDD, we are confident that this threat did not materialize. 

Additionally, our experiment was exposed to the attrition threat (loss of two participants). 

\textbf{Construct validity} refers to the correctness in the mapping between the theoretical constructs to be investigated and the operationalizations of the variables in the study.

The study suffers from the mono-method bias threat since only one metric was used to measure the response variable (i.e., quality). This issue was mitigated by interpreting the results jointly with BE and RE1 (see below).

The concepts underlying the cause construct used in the experiment appear to be clear enough to not constitute a threat. The TDD cycle was explained according to the literature \cite{beck2003test}. However, some articles point out that TDD is a complex process and might not be consistently applied by developers \cite{aniche2010most}. Conformance to the development approaches is one of the big threats to construct validity that might have materialized in this and most (if not all) other experiments on TDD. However, we tried to minimize this threat to validity by supervising the students while they coded, and encouraging them to adhere as closely as possible to the development approaches taught during the laboratory.

There are no significant social threats, such as evaluation apprehension: all subjects participated on a voluntary basis in the experiment and were free to drop out of the sessions if they so wished.

\textbf{External validity} relates to the possibility of generalizing the study results beyond the study's objects and subjects \cite{wohlin2012experimentation}.

The experiment was exposed to the selection threat since we could not randomly select participants from a population; instead, we had to rely on convenience sampling. Convenience sampling is an endemic threat in SE experiments. This issue was taken into account when reporting the results, acknowledging that the findings are only valid for developers with no previous experience in TDD.  

Java was used as the programming language for the experimental sessions and measuring the response variable with acceptance test suites. This way, we addressed possible threats regarding the use of different programming languages to measure the response variable. However, this limits the validity of our results to this language only.

Three out of the four tasks (MR, BSK, and SDK) used in the experiment were toy greenfield tasks. This affects the generalizability of the results and their applicability in industrial settings. The task domain might not be representative of real-life applications, and the duration of the experiment (two hours and 15 minutes to perform each task) might have had an impact on the results. We acknowledge that this might be an obstacle to the generalizability of the results outside the artificial setting of a laboratory. We take this into account when reporting our findings, as they are only valid for toy tasks.

We acknowledge as a threat to validity the use of students as subjects: however, this threat was minimized as they are graduate students close to the end of their educational programmes. Even so, this still limits the generalization of our results beyond novice developers.

\section{Comparison of Results Across Replications}
\label{ten}

Table \ref{results_tdd_itl_replications} shows the results achieved across the replications for the research question---both in terms of statistical significance and effect size. 

\begin{table}[h!]
\caption{Results for TDD vs. ITL across replications: statistical significance and effect size.}
\centering
\label{results_tdd_itl_replications}
\begin{tabular}{l|l|l|l} \hline
 & \textbf{BE} & \textbf{RE1} & \textbf{RE2} \\ \hline\hline
\textbf{Significance}  & ITL=TDD & ITL=TDD & ITL=TDD  \\ \hline
\textbf{Effect size} & ITL$>$TDD (BSK) & ITL$<$TDD (BSK) & ITL$<$TDD (BSK, MR)  \\ 
& &  & ITL$>$TDD (MF, SDK)  \\ \hline
\end{tabular}
\end{table}

As Table \ref{results_tdd_itl_replications} shows, RE1's and RE2's results are consistent for BSK (ITL$<$TDD). However, such results are inconsistent with BE's  (ITL$>$TDD). This may be due to differences between BE's settings and both RE1's and RE2's common settings. Let us focus on the differences among BE, RE1 and RE2 and their settings, and discuss whether such differences may have impacted the results (i.e., let us look for moderator variables). 

The directionality of the statistical test in BE and RE1 seems not to have affected the results. This is because the difference in effectiveness between TDD and ITL was not statistically significant in any replication. This suggests that the threat to conclusion validity did not materialize neither in BE nor RE1.

We hypothesized that the participants applying ITL in BE and RE1 might be less motivated than those applying TDD---perhaps due to ITL's lower desirability in an experiment on TDD. This may have distorted results. RE2 overcame such shortcoming as we made the participants apply both development approaches twice. RE2's results were similar to RE1's. However, as BE's results are the opposite, we cannot assess whether the disagreement comes from the materialization of such threat, or because other confounding factors. This could be studied in posterior replications.

In RE1 there were two confounding factors that might have affected the results: pair-programming (as some subjects were paired due to space restrictions), and the participation of a mixture of graduate and undergraduate students. After eliminating such confounding factors in RE2, we noticed that RE2's results were similar to RE1's. However, we cannot ascertain whether such confounding factors affected RE1's results. This is because RE1's results do not match BE's---where no confounding is in place either. We pose the disagreement between BE's and RE2's results to the different populations in the experiments---undergraduates in BE and graduates in RE2. However this is only an hypothesis that needs further studying.

While BE's participants could work inside and outside the laboratory, RE1's and RE2's participants were only allowed to work inside the laboratory. The differences in results observed across BE and both RE1 and RE2 suggest that this threat to validity may have materialized in BE. This suggests that TDD seems to outperform ITL in controlled environments, while TDD's performance tends to deteriorate to a larger extent than ITL's in uncontrolled environments. 
The stubs provided in RE1 and RE2 may have had a positive impact on the quality achieved with TDD. We make such observation because TDD outperforms ITL whenever stubs are provided---and viceversa when not, as in BE. However, we should be cautious making such interpretation. This is because the experimental session length also varied along the provision of stubs. Besides, the influence of such time reduction could not be assessed because the experimental session length was changed across all the replications. Thus, no combination of replications allowed assessing the stability of the findings after fixing session length. Again, the impact of session length on results shall be assessed in subsequent replications. 

BE's participants were assigned to either the ITL or TDD groups based on an ad-hoc skill set. RE1 overcame this threat to validity by assigning the participants to the treatments by means of full-randomization. RE2 by making all the participants apply both development approaches twice. As BE's results and those of RE1 and RE2 do not match, this may suggest BE's participant assignment based on skills may have had an impact on results. 

We noticed that leakage was possible from one treatment to another in RE1. In particular, RE1's participants might apply a mixed development approach in the experiment---as they were trained in both development approaches before the experimental session. This was solved in BE and in RE2 by only training the participants in the development approach to be applied in the immediate experimental session. However, while RE2's results match RE1's, they disagree with BE's results. Such contradicting results may point towards the presence of other confounding variables impacting the results.

RE2's within-subjects design materialized an additional threat to validity: carry-over (i.e., the impact of the application of one treatment over another). In particular, as the participants already applied ITL before TDD, this may have boosted TDD's effectiveness. Even though RE2's results match RE1's---where carry-over is not possible---both disagree with BE's---where carry-over is neither possible. Again, such contradicting results shall be further investigated in posterior replications.

Despite RE1 used an arbitrary threshold for considering a user story as delivered (i.e., 50 per cent of assert statements passed), RE1's results and RE2's agree. However, we do not know whether the variation of such element impacted the results---as the 50 per cent threshold was also used in BE, and contradictory results were reached with regards to RE1. This issue shall be investigated in posterior replications.

BE's user story weights may have had an impact on results because RE1's and RE2's results match (and they do not use such weights for measuring QLTY), and they are contrary to BE's. 

Mono-operation bias materialized in BE and RE1 because the participants only coded BSK. RE2's different tasks (MR, MF and SDK) made it possible studying TDD's and ITL's performance under a larger array of coding problems. RE2's results suggested that the direction and magnitude of the effect size depend upon the task. In other words, some tasks favour TDD (MR and BSK) and others ITL (MF and SDK). This was not previously noticed in BE and RE1, and adds to the body of evidence. 

Table \ref{elements_impacting} shows the threats to validity of the replications---grouped by the dimension to which they belong---and an assessment of their possible impact on results. As we can see, five different sources of variability may have impacted the replications' results: the allowance to work outside the laboratory in BE, the allocation of subjects to treatments (assignment attending to skills in BE), the lack of stubs in BE, the weighting element used in the construct for QLTY in BE and mono-operation bias in BE and RE1. The directionality of the statistical tests seem not to have any impact on the results, while the impact of the rest of the elements could not be assessed.

\begin{table}[h!]
\small
\caption{Elements potentially influencing results across the replications.}
\label{elements_impacting}
\begin{center}
\begin{tabular}{l|l|l|l} \hline
    \textbf{Impact} & \textbf{Element} & \textbf{Source} & \textbf{Threat to validity}  \\ \hline \hline
    No & Directionality Statistical tests & BE, RE1 & Conclusion validity \\ \cline{1-4} 
    - & Compensatory rivalry bias & BE, RE1 & \multirow{8}{*}{Internal validity} \\ \cline{1-3}
    - & Confounding: pair programming and subject type & RE1 & \\ \cline{1-3}
    Yes & Experimental Task Execution & BE & \\ \cline{1-3}
    Yes & Stubs Provision & RE1, RE2 & \\  \cline{1-3}     
    - & Experimental Task Duration & BE, RE1, RE2 &   \\ \cline{1-3}   
    Yes & Subjects to Treatment Allocation & BE & \\ \cline{1-3}
    - & Training leakeage & RE1 & \\ \cline{1-3}    
    - & Carry-over & RE2 & \\ \cline{1-4}    
    - & QLTY operationalization (50\% threshold) &BE, RE1 & \multirow{2}{*}{Construct validity}\\ \cline{1-3}    
    Yes & QLTY operationalization (Weighting) & BE & \\ \cline{1-4}  
    Yes & Mono-Operation Bias & BE, RE1 & \multirow{2}{*}{External validity} \\ \cline{1-3} 
    - & Population & BE, RE1 & \\ \hline
\end{tabular}
\end{center}
\end{table}

\section{Conclusions and Future Work}
\label{six}

Experiments on TDD tend to provide conflicting results (i.e., positive, negative, and neutral) in terms of external quality. This could be due to the many dimensions in which the experiments vary (e.g., experimental design, artifacts, context variables, etc.). A systematic approach towards replication may help to stabilize the experiments' results, and facilitate the discovery of previously unknown moderator variables.

We run a replication of Erdogmus et al.'s experiment on TDD \cite{fucci2013replicated}. Such experiment was already replicated by Fucci et al. \cite{fucci2013replicated}. We tweaked our replication's design to overcome the threats to validity of the previous experiments. We also purposely varied our replication's tasks to increase the external validity of results (i.e., by making the participants code four, instead of one task). This strengthened the evidence obtained, and allowed us to see that the allowance to work outside the laboratory, the provision of stubs, the allocation of subjects to treatments, the operationalization of the response variable, and the task being developed may be affecting TDD experiments' results on software quality. 

As others did before \cite{offutt2018don}, we propose as a further line of research to conduct replications studying other response variables rather than external quality (e.g., internal quality, maintainability, etc.). We also propose to study which tasks' characteristics make them more suitable to be developed with TDD. This may assist practitioners when choosing development methods to create new pieces of software. 

In sum, by means of an illustrative example we showed how replications allow increasing the reliability of the findings and hypothesizing on moderator variables. By reflecting upon previous experiment's limitations it is possible tweaking replications' designs and overcoming previous experiments' weaknesses. This may aid to strengthen the evidence of the results, and to uncover yet to explore lines of research. 


\bibliographystyle{spmpsci}      
\bibliography{biblio}   

\end{document}


\title{Increasing Validity Through Replication: An Illustrative TDD Case}

\author{Adrian Santos \and Sira Vegas \and Fernando Uyaguari \and Oscar Dieste \and Burak Turhan \and Natalia Juristo 
}


\institute{Adrian Santos \at
              M3S-ITEE University of Oulu, Finland \\
              \email{adrian.santos.parrilla@oulu.fi}           
           \and
           Sira Vegas, Oscar Dieste, Natalia Juristo \at
              Escuela T\'ecnica Superior de Ingenieros Inform\'aticos, Universidad Polit\'ecnica de Madrid, Spain \\
              \email{svegas/odieste/natalia@fi.upm.es} 
            \and 
            Fernando Uyaguari \at
            ETAPA, Ecuador\\
            \email{fuyaguari01@gmail.com} 
            \and
            Burak Turhan \at Department of Computer Science, Brunel University London, England \\
            \email{burak.turhan@brunel.ac.uk}
}

\date{Received: 01/03/18 / Accepted: -}

\maketitle

\section*{Appendix A: BSK}
\begin{scriptsize}

The objective is to develop an application that can calculate the score of a single bowling  game using TDD. There is no graphical user interface. You will only work with objects and JUnit test cases in this assignment. You will not need a main method.
The application's requirements are divided into a set of user stories, which serve as your to-do list. You should be able to incrementally develop a complete solution without an upfront comprehension of all the game's rules. Do not read ahead, and handle the requirements one at a time in the order provided. Solve the problem using TDD, starting with the first story's requirement. Remember to always lead with a test case, taking hints from the examples provided. Only when you are done with a story, move on to the next one. A story is done when you are confident that your program correctly implements all the functionality stipulated by the story's requirement, meaning all of your test cases for that story and all of the test cases for the previous stories pass. You may need to tweak your solution as you progress towards more advanced requirements.
\\ 
\textbf{1. Frame}	
Each turn of a bowling game is called a frame. Ten pins are arranged in each frame. The goal of the player is to knock down as many pins as possible in each frame. The player has two chances, or throws, to do so. The value of a throw is given by the number of pins knocked down in that throw.

\textbf{Requirement}: Define a frame as composed of two throws. The first and second throws should be distinguishable.

\textbf{Example}: [2, 4] is a frame with two throws, in which two pins were knocked down in the first throw and four pins were knocked down in the second.
\\
\textbf{2. Frame Score}
An ordinary frame's score is the sum of its throws.

\textbf{Requirement}: Compute the score of an ordinary frame.

\textbf{Examples}: The score of the frame [2, 6] is 8. The score of the frame [0, 9] is 9.
\\
\textbf{3. Game}
A single game consists of 10 frames.

\textbf{Requirement}: Define a game, which consists of 10 frames.

\textbf{Example}: The sequence of frames [1, 5] [3, 6] [7, 2] [3, 6] [4, 4] [5, 3] [3, 3] [4, 5] [8, 1] [2, 6] represents a game. You will reuse this game from now on to represent different scenarios, modifying only a few frames each time.
\\
\textbf{4. Game Score}
The score of a bowling game is the sum of the individual scores of its frames.

\textbf{Requirement}: Compute the score of a game.

\textbf{Example}: The score of the game [1, 5] [3, 6] [7, 2] [3, 6] [4, 4] [5, 3] [3, 3] [4, 5] [8, 1] [2, 6] is 81.
\\
\textbf{5. Strike}
A frame is called a strike if all 10 pins are knocked down in the first throw. In this case, there is no second throw. A strike frame can be written as [10, 0]. The score of a strike equals 10 plus the sum of the next two throws of the subsequent frame.

\textbf{Requirement}: Recognize a strike frame. Compute the score of a strike. Compute the score of a game containing a strike.

\textbf{Examples}: Suppose [10, 0] and [3, 6] are consecutive frames. Then the first frame is a strike and its score equals 10 + 3 + 6 = 19. The game [10, 0] [3, 6] [7, 2] [3, 6] [4, 4] [5, 3] [3, 3] [4, 5] [8, 1] [2, 6] has a score of 94.
\\
\textbf{6. Spare}
A frame is called a spare when all 10 pins are knocked down in two throws. The score of a spare frame is 10 plus the value of the first throw from the subsequent frame.

\textbf{Requirement}: Recognize a spare frame. Compute the score of a spare. Compute the score of a game containing a spare frame.

\textbf{Examples}: [1, 9], [4, 6], [7, 3] are all spares. If you have two frames [1, 9] and [3, 6] in a row, the spare frame's score is 10 + 3 = 13. The game [1, 9] [3, 6] [7, 2] [3, 6] [4, 4] [5, 3] [3, 3] [4, 5] [8, 1] [2, 6] has a score of 88.
\\
\textbf{7. Strike and Spare}
A strike can be followed by a spare. The strike's score is not affected when this happens.

\textbf{Requirement}: Compute the score of a strike when it is followed by a spare. Compute the score of a game with a spare following a strike.

\textbf{Examples}: In the sequence [10, 0] [4, 6] [7, 2], a strike is followed by a spare. In this case, the score of the strike is 10 + 4 + 6 = 20, and the score of the spare is 4 + 6 + 7 = 17. The game [10, 0] [4, 6] [7, 2] [3, 6] [4, 4] [5, 3] [3, 3] [4, 5] [8, 1] [2, 6] has a score of 103.
\\
\textbf{8. Multiple Strikes}
Two strikes in a row are possible. You must take care when this happens for the computation of the first strike's score requires the values of throws from two subsequent frames.

\textbf{Requirement}: Compute the score of a strike that is followed by another strike. Compute the score of a game with two strikes in a row.

\textbf{Examples}: In the sequence [10, 0] [10, 0] [7, 2], the score of the first strike is 10 + 10 + 7 = 27. The score of the second strike is 10 + 7 + 2 = 19. The game [10, 0] [10, 0] [7, 2] [3, 6] [4, 4] [5, 3] [3, 3] [4, 5] [8, 1] [2, 6] has a score of 112.
\\
\textbf{9. Multiple Spares}
Two spares in a row are possible. The first spare's score is not affected when this happens.

\textbf{Requirement}: Compute the score of a game with two spares in a row.

\textbf{Example}: The game [8, 2] [5, 5] [7, 2] [3, 6] [4, 4] [5, 3] [3, 3] [4, 5] [8, 1] [2, 6] has a score of 98.
\\
\textbf{10. Spare as the Last Frame}
When a game's last frame is a spare, the player will be given a bonus throw. However, this bonus throw does not belong to a regular frame. It is only used to calculate the score of the last spare.

\textbf{Requirement}: Compute the score of a spare when it is the last frame of a game. Compute the score of
a game when its last frame is a spare.

\textbf{Example}: The last frame in the game [1, 5] [3, 6] [7, 2] [3, 6] [4, 4] [5, 3] [3, 3] [4, 5] [8, 1] [2, 8] is a spare. If the bonus throw is [7], the last frame has a score of 2 + 8 + 7 = 17. The game has a score of 90.
\\
\textbf{11. Strike as the Last Frame}
When a game's last frame is a strike, the player will be given two bonus throws. However, these two bonus throws do not belong to a regular frame. They are only used to calculate the score of the last strike frame.

\textbf{Requirement}: Compute the score of a spare when it is the last frame of a game. Compute the score of a game when the last frame is a strike.

\textbf{Example}: The last frame in the game [1, 5] [3, 6] [7, 2] [3, 6] [4, 4] [5, 3] [3, 3] [4, 5] [8, 1] [10, 0] is a strike. If the bonus throws are [7, 2], the last frame's score is 10 + 7 + 2 = 19. The game's score is 92.
\\
\textbf{12. Bonus is a Strike}
Further bonus throws are not granted when a game's last frame is a spare and the bonus throw is a strike.

\textbf{Requirement}: Compute the score of a game in which the last frame is a spare and the bonus throw is a strike.

\textbf{Example}: In the game [1, 5] [3, 6] [7, 2] [3, 6] [4, 4] [5, 3] [3, 3] [4, 5] [8, 1] [2, 8], the last frame is a spare. If the bonus throw is [10], the game's score is 93.
\\
\textbf{13. Best Score}
A perfect game consists of all strikes (a total of 12, including the bonus throws), and has a score of 300.

\textbf{Requirement}: Check that the score of a perfect game is 300.

\textbf{Example}: A perfect game looks like [10, 0] [10, 0] [10, 0] [10, 0] [10, 0] [10, 0] [10, 0] [10, 0] [10,0] [10, 0] with bonus throws [10, 10]. It's score is 300.
\\
\textbf{14. Real Game}

\textbf{Requirement}: Check that the score of the game [6, 3] [7, 1] [8, 2] [7, 2] [10, 0] [6, 2] [7, 3] [10, 0] [8, 0] [7, 3] [10] is 135.

Congratulations, you are done!
\end{scriptsize}
\section*{Appendix B: Mars Rover API}
\begin{scriptsize}
\begin{itemize}
    \item Develop an API that moves a rover around on a grid.
    \item You are given the initial starting point (x,y) of a rover and the direction (N,S,E,W) in which it is facing.
    \item The rover receives a character array of commands
    \item Implement commands that move the rover forward/backward (f,b)
    \item Implement commands that turn the rover left/right (l,r)
    \item Implement wrapping from one edge of the grid to another (planets are spheres after all)
    \item Implement obstacle detection before each move to a new square. If a given sequence of commands encounters an obstacle, the rover moves up to the last possible point and reports the obstacle.
    \item Example: The rover is on a 100x100 grid at location (0, 0) and facing NORTH. The rover is given the commands “ffrff” and should end up at (2,2).
\end{itemize}
\end{scriptsize}

\section*{Appendix C: Sudoku}
\begin{scriptsize}
Sudoku is a game with a few simple rules, where the goal is to place nine sets of positive digits (1…9) into the cells of a fixed grid structure. A valid Sudoku solution should conform to the following rules:
\begin{itemize}
    \item A cell in a Sudoku game can only store positive digits, i.e. 1...9.
    \item A "sub-grid" is a 3x3 arrangement of cells.
    \item All digits appear only once in a sub-grid, i.e., they cannot be repeated.
    \item The Sudoku board (or global grid) consists of a 3x3 arrangement of sub-grids, yielding a 9x9 arrangement of cells.
    \item A digit can appear only once in the rows of the global grid.
    \item A digit can appear only once in the columns of the global grid.
\end{itemize}

\textbf{Your task is to check the validity of a given solution for a Sudoku game}:
	
\begin{enumerate}
    \item You should read the candidate solution from a string variable, which would have been displayed as below, when printed on the screen:
\begin{table}[ht]
\tiny
\begin{center}
\label{settings_baseline_replications}
\begin{tabular}{ p{0.05cm}p{0.05cm}p{0.05cm}p{0.05cm}p{0.05cm}p{0.05cm}p{0.05cm}p{0.05cm}p{0.05cm} }
1&	2&	3&	4&	5&	6&	7&	8&	9 \\
9&	1&	2&	3&	4&	5&	6&	7&	8 \\
8&	9&	1&	2&	3&	4&	5&	6&	7 \\
7&	8&	9&	1&	2&	3&	4&	5&	6 \\
6&	7&	8&	9&	1&	2&	3&	4&	5 \\
5&	6&	7&	8&	9&	1&	2&	3&	4 \\
4&	5&	6&	7&	8&	9&	1&	2&	3 \\
3&	4&	5&	6&	7&	8&	9&	1&	2 \\
2&	3&	4&	5&	6&	7&	8&	9&	1 \\
\end{tabular}
\end{center}
\end{table}
    \item Check whether the provided string follows the correct format (i.e., 9 lines with 9 entries in each line).
    \item Check the validity of the candidate solution against the rules listed above.
    \item Throw "CustomSudokuException" for all error cases including but not limited to: wrong format for a solution string.
    \item Implement the functionality in your program to return a string message on the validity of the solution:
    \begin{itemize}
        \item If it is valid, the following message shall be displayed: "This is a valid solution", followed by the solution itself (as above).
        \item If the solution is not valid, you shall return a failure message, indicating the reason why it is not valid, e.g., "9 appears more than once in row 5" (Assume that the lower left corner of the grid has the coordinates (1, 1)). Again, you shall display the solution after this error message.
    \end{itemize}
\end{enumerate}
\end{scriptsize}

\section*{Appendix D: Music Fone}

\begin{scriptsize}

MusicPhone is an application that runs on a GPS-enabled, MP3-capable mobile phone. MusicPhone will make recommendations for artists that the user may like and find upcoming concert events for artists using data gathered from the Last.FM website. The goal of the following tasks is to implement the necessary logic.
\\
\\
\textbf{Task A: Ramp-up}

\textbf{A1}. Run the project (Select the project, right click and select Run As -$>$ Run Configurations. From the configurations, select MusicPhone from C/C++ Application) and see three UI windows appear. The Player and GPS UIs are complete. The Recommender window is just a skeleton. You will implement most of the required functionality in the Recommender class. 
		
\textbf{A2}. Run the project with Unit Test configuration (Select the project, right click and Run As-$>$Run Configurations. From the configurations, select MusicPhone UnitTest from C/C++ Unit) and view the green bar. Check the structure of the sample test in SmokeTest.cpp to get familiar with the application. 

\textbf{A3}. Take 5-10 minutes to read through the information in the provided documentation.
\\
\textbf{Task B: Compute distance to a concert}

The user will see how far away upcoming concert events are for a particular artist based on the user's current GPS position and the position of the concert venue. Implement the Distance class in \texttt{commons.dataClasses} and a public method with the signature \texttt{public static double computeDistance (GeoPoint, GeoPoint, String)} in this class to calculate the great-circle distance between two geo-coded positions. A geo-coded position is represented by a \texttt{GeoPoint} object that specifies a latitude and longitude in degrees. The method computes the great-circle distance between two points in either kilometres ("km") or miles ("mi") as specified by the third parameter. This parameter can be lower case, upper case or a combination of both. Valid latitude values are between -90 and 90 degrees; valid longitude values are between -180 and 180 degrees. Sometimes a \texttt{GeoPoint} has an invalid latitude or longitude. In these cases, the distance returned should be -1. 

\begin{table}[!htbp]
\tiny
\centering
\label{my-label}
\begin{tabular}{|c|c|c|c|c|c|}
\hline
\multicolumn{2}{|c|}{Geopoint_{1}}                                                                                  & \multicolumn{2}{c|}{Geopoint_{2}}                                                                                   & \multirow{2}{*}{\textit{units}} & \multirow{2}{*}{\begin{tabular}[c]{@{}c@{}}\textbf{Great-circle}\\ \textbf{distance} (in \textit{units})\end{tabular}} \\ \cline{1-4}
\begin{tabular}[c]{@{}c@{}}LatR_{1}\\ LatD_{1}\end{tabular}  & \begin{tabular}[c]{@{}c@{}}LonR_{1}\\ LonD_{1}\end{tabular}   & \begin{tabular}[c]{@{}c@{}}LatR_{2}\\ LatD_{2}\end{tabular} & \begin{tabular}[c]{@{}c@{}}LonR_{2}\\ LonD_{2}\end{tabular}    &                        &                                                                                             \\ \hline
\begin{tabular}[c]{@{}c@{}}\textit{0}\\ 0\end{tabular} & \begin{tabular}[c]{@{}c@{}}\textit{0}\\ 0\end{tabular}           & \begin{tabular}[c]{@{}c@{}}\textit{0}\\ 0\end{tabular}         & \begin{tabular}[c]{@{}c@{}}\textit{0}\\ 0\end{tabular}            & km                     & 0                                                                                           \\ \hline
\begin{tabular}[c]{@{}c@{}}\textit{0}\\ 0\end{tabular}          & \begin{tabular}[c]{@{}c@{}}\textit{0}\\ 0\end{tabular}           & \begin{tabular}[c]{@{}c@{}}\textit{1.047}\\ 60\end{tabular}    & \begin{tabular}[c]{@{}c@{}}\textit{0}\\ 0\end{tabular}            & km                     & 6671.70                                                                                     \\ \hline
\begin{tabular}[c]{@{}c@{}}\textit{0}\\ 0\end{tabular}          & \begin{tabular}[c]{@{}c@{}}\textit{0}\\ 0\end{tabular}           & \begin{tabular}[c]{@{}c@{}}\textit{6.283}\\ 360\end{tabular}   & \begin{tabular}[c]{@{}c@{}}\textit{6.283}\\ 360\end{tabular}      & mi                     & -1                                                                                          \\ \hline
\begin{tabular}[c]{@{}c@{}}\textit{0}\\ 0\end{tabular}          & \begin{tabular}[c]{@{}c@{}}\textit{0}\\ 0\end{tabular}           & \begin{tabular}[c]{@{}c@{}}\textit{1.047}\\ 60\end{tabular}    & \begin{tabular}[c]{@{}c@{}}\textit{0}\\ 0\end{tabular}            & mi                     & 4145.60                                                                                     \\ \hline
\begin{tabular}[c]{@{}c@{}}\textit{0.630}\\ 36.12\end{tabular}  & \begin{tabular}[c]{@{}c@{}}\textit{-1.513}\\ -86.67\end{tabular} & \begin{tabular}[c]{@{}c@{}}\textit{0.592}\\ 33.94\end{tabular} & \begin{tabular}[c]{@{}c@{}}\textit{-2.066}\\ -118.40\end{tabular} &   mi                     & 1793.55                                                                                     \\ \hline

\multicolumn{6}{r}{\begin{tabular}[c]{@{}r@{}}Numbers in roman type are in degrees.\\ \textit{Numbers in italics are in radians}\end{tabular}}
\end{tabular}
\end{table}
The formula for the great-circle distance is:
\\

\begin{scriptsize}lonD\end{scriptsize}= longitude in degrees

\begin{scriptsize}latD\end{scriptsize}= latitude in degrees

\begin{scriptsize}\textit{lonR}\end{scriptsize}= longitude in radians

\begin{scriptsize}\textit{latR}\end{scriptsize}= latitude in radians

\begin{scriptsize}\textit{radians=(degrees x $\pi$)/100}\end{scriptsize}

$a = \sqrt{\sin^2(\frac{\Delta LatR}{2})+ \cos({LatR_{1}}) x \cos({LatR_{2}}) x \sin^2(\frac{\Delta LonR}{2})}$

\begin{scriptsize}Great circle distance = $2 x \arcsin(min(1.0,a))x Radius$ \end{scriptsize}

where \textit{Radius} is the Earth's radius in kilometres (6371.01) or in miles (3958.76):
\\

\\
\textbf{Task C: Find concerts for an artist} 

Implement the \texttt{getDestinationsForArtist} method of the \texttt{Recommender} class. When the user clicks on an artist's name in the list of recommended artists, the UI calls the \texttt{getDestinationsForArtist} method to obtain a list of destinations, where each destination contains the information on an upcoming concert together with the distance to the concert's location from the user's current position. The user's current position is provided by the \texttt{GPS} class. If the concert's location has an invalid \texttt{GeoPoint}, then the distance should be marked as -1. You will need to use \texttt{IConnector's GetConcertForArtist} method to implement this task. 
\\
\textbf{Task D: Recommend artists}

Implement the \texttt{getRecommendations} method of the \texttt{Recommender} class. MusicPhone recommends artists based on the favourite artists of Last.FM users who are the top fans of the artist currently playing on the user's player. MusicPhone recommends the 20 most-frequently-appearing artists among the fans' top artists (as in "people who like this artist also like these other artists..."). \texttt{getRecommendations} should return a list of \texttt{Recommendation} objects. Each \texttt{Recommendation} object consists of an artist's name (artist) and the frequency of occurrence (\texttt{fanCount}) of that artist. You will need to use \texttt{IConnector's getTopFansForArtist} and \texttt{getTopArtistsByFan} methods to implement this method. 
\textit{Example:} Suppose the currently playing artist is \textit{Cher}. Suppose \textit{John}, \textit{Sally}, and \textit{Tom} are the only top fans of Cher. If \textit{Tom} and \textit{John} both like \textit{U2} as well, then \texttt{getRecommendations} should return a list that includes a recommendation containing \textit{U2} with a \texttt{FanCount} of 2.
\\
\textbf{Task E: Compute an itinerary for concerts}

\textbf{E0.} Implement the \texttt{buildItineraryForArtists} method of the \texttt{Recommender} class. In the UI, the user can click on buttons that will add (Add button) or remove (Remove button) a selected artist to a list of artists who the user would like to see in concert. Whenever this list is changed, the application recalculates an itinerary of upcoming concerts by calling this method. An itinerary is a list of \texttt{Destination} objects, where each object contains the concert information and the distance to the concert's location from the previous destination. The first destination's distance in the list is the distance from the user's current position. The itinerary must be chronologically ordered according to the start date of the concerts that it includes. Address and satisfy the constraints \textit{one by one} in any order: 

\textbf{E1.} The itinerary starts with the concert having the earliest start date for any of the selected artists. 

\textbf{E2.} An artist has at most one concert in the itinerary (some artists may have no concerts, however). 

\textbf{E3.} No two concerts in the itinerary may take place on the same day. 

\textbf{E4.} If multiple artists have concerts on the same day, the itinerary includes only the concert that is closest in distance to the previous destination on the list. 

\textbf{E5.} If the same artist has multiple concerts on the same day, the itinerary includes only the concert that is closest in distance to the previous destination on the list.
\\
\\

\begin{normalsize}\textbf{MusicPhone Project layout}\end{normalsize} \\

\underline{\textbf{app}} - Package containing the UI resources and classes that provide data from Last.FM. \\

\underline{\textbf{commons}} - Package containing the data models, interfaces and logic modules used by the MusicPhone UI and tests. \textit{You will implement the missing logic inside this project}. You may \textbf{not} change any of the interfaces or the \texttt{DeviceManager}. You may extend the behaviour of the \texttt{Recommender}, but without changing its existing behaviour or its interface. You may add one or more classes here if required. \\

\underline{\textbf{commons.dataClasses}} - Package containing the basic classes representing the entitites present in MusicPhone. \\

\underline{\textbf{commons.interfaces}} - Package containing the interfaces that are implemented by the different parts of this system. You should not change any of these. \\

\underline{\textbf{commons.xmlData}} - Package containing the dump of Last.FM response for the API necssary to solve the task in XML format. These are useful for testing purposes. \\

\underline{\textbf{commons.dataConnectors}} - Package containing the implementation of IConnector (LastFmXmlConnector) methods to parse the Last FM XML files. You may want to instantiate this class for testing purposes. \\

\underline{\textbf{gps}}, \underline{\textbf{player}}, and \underline{\textbf{recommender}} - Packages containing the UIs for the application main components. The UI is bounded to a class (e.g., GpsUI.java and Gps.java) which is the implementation of the interfaces present in commons.interfaces. \\
\\
\begin{normalsize}\textbf{What you should know before you start}\end{normalsize}
\\ \\
\textbf{Initialization of components}

The application project creates specific instances of \texttt{IPlayer}, \texttt{IGps}, and \texttt{IRecommender} objects. These instances persist when the application is running. The application will set the \texttt{Connector} property of the \texttt{Recommender} object to an instance of \texttt{LastFmXmlConnector}, which implements the \texttt{IConnector} interface.
\\
\textbf{The IConnector interface}

Defines access to XML data from Last.FM. The \texttt{IConnector} class you need to test your implementation with preloaded XML data is \texttt{LastFmXmlConnector}. This class has a 0-argument constructor. To access the XML data from \texttt{\textbf{Commons}}, use the \texttt{Connector} property of the \texttt{Recommender class}.
\\
\textbf{XML data for testing}

Located in the XmlData folder of the \textbf{Commons} project. The \texttt{LastFmXmlConnector} class accesses the files in this folder.
\\
\textbf{DeviceManager.Instance}

Provides singleton access to instances of the \texttt{IPlayer} and \texttt{IGps} objects. When these objects are instantiated by the application, they register with the \texttt{DeviceManager}.
\\
\textbf{Architecture}

Review the block diagram on the next page.

\begin{figure}[h!]
\centering
  \includegraphics[height=5cm]{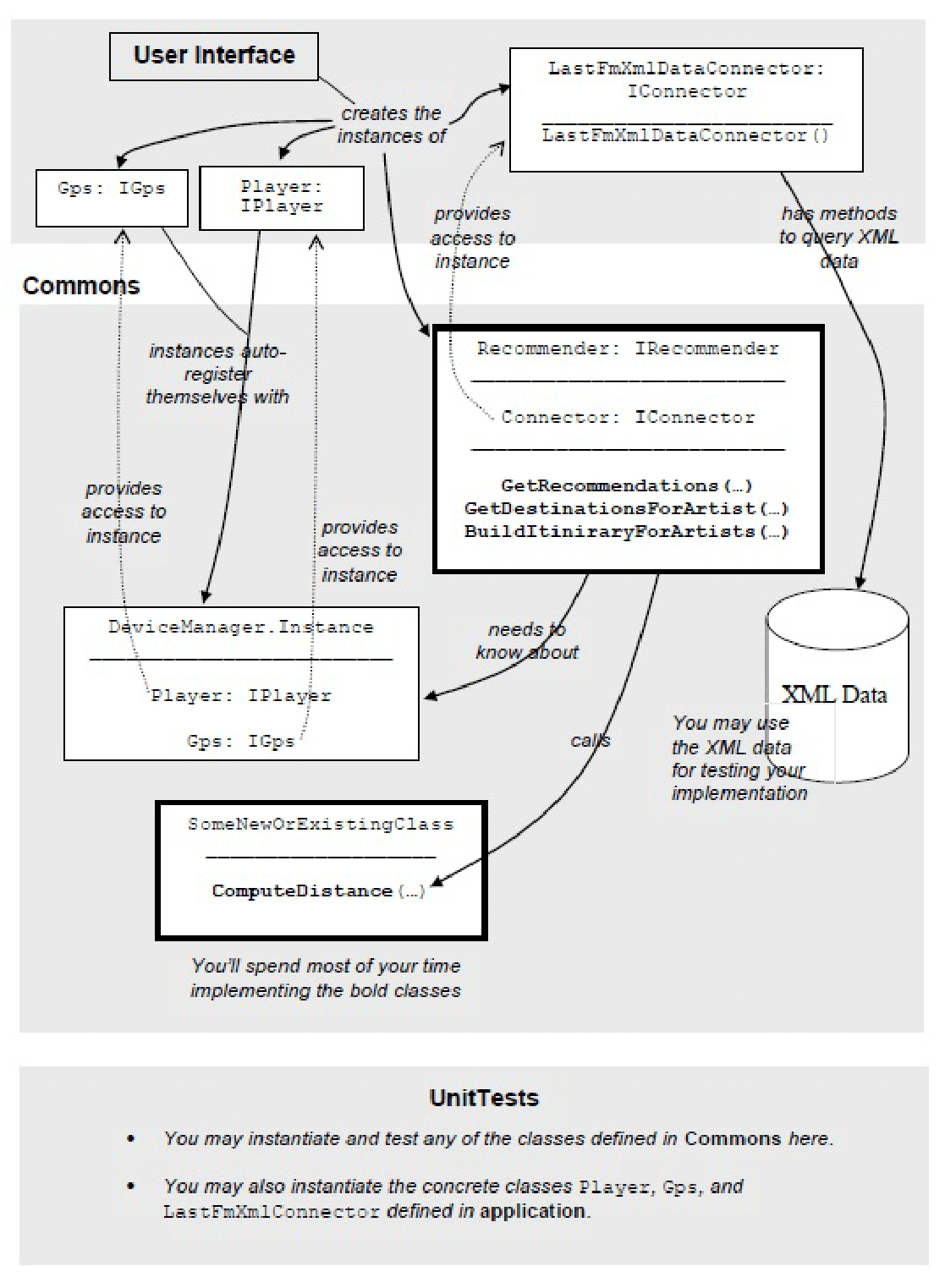}
\end{figure}

\end{scriptsize}
%

\section*{Appendix E: Residuals LMM for QLTY}

\begin{figure}[h!]
\centering
  \includegraphics[height=8cm]{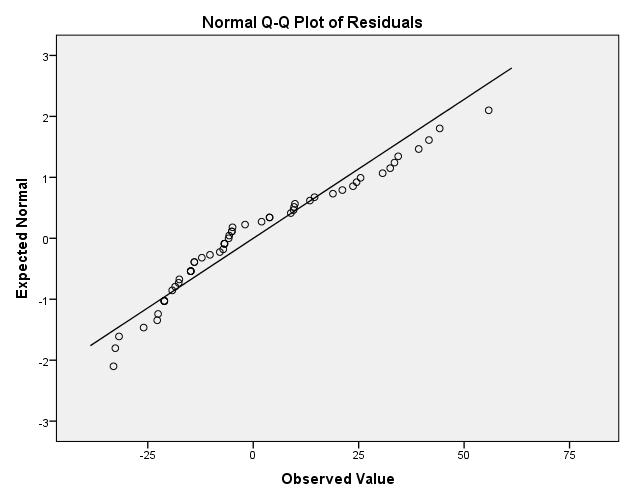}
\end{figure}